\documentclass[epsfig]{article}
\usepackage{epsfig}
\usepackage{amsmath}
\newcommand{\ba}{\begin{eqnarray}}
\newcommand{\ea}{\end{eqnarray}}
\newcommand{\be}{\begin{equation}}
\newcommand{\ee}{\end{equation}}
\catcode`\@=11
\catcode`\@=11

\def\section{\@startsection{section}{1}{\z@}{3.5ex plus 1ex minus
   .2ex}{2.3ex plus .2ex}{\large\bf}}

\def\ps@headings{\def\@oddfoot{}\def\@evenfoot{}
\def\@oddhead{\hbox{}\hfill
        \makebox[.5\textwidth]{\raggedright\ignorespaces --\thepage{}--
        \hfill }}
\def\@evenhead{\@oddhead}
\def\subsectionmark##1{\markboth{##1}{}}
}

\ps@headings

\catcode`\@=12

\relax

\def\figcap{\section*{Figure Captions\markboth
        {FIGURECAPTIONS}{FIGURECAPTIONS}}\list
        {Fig. \arabic{enumi}:\hfill}{\settowidth\labelwidth{Fig. 999:}
        \leftmargin\labelwidth
        \advance\leftmargin\labelsep\usecounter{enumi}}}
 \relax
\def\tablecap{\section*{Table Captions\markboth
        {TABLECAPTIONS}{TABLECAPTIONS}}\list
        {Table \arabic{enumi}:\hfill}{\settowidth\labelwidth{Table 999:}
        \leftmargin\labelwidth
        \advance\leftmargin\labelsep\usecounter{enumi}}}
 \relax
\def\reflist{\section*{References\markboth
        {REFLIST}{REFLIST}}\list
        {[\arabic{enumi}]\hfill}{\settowidth\labelwidth{[999]}
        \leftmargin\labelwidth
        \advance\leftmargin\labelsep\usecounter{enumi}}}
 \relax

\catcode`\@=11

\def\marginnote#1{}
\newcount\hour
\newcount\minute
\newtoks\amorpm
\hour=\time\divide\hour by60 \minute=\time{\multiply\hour by60
\global\advance\minute by- \hour}
\edef\standardtime{{\ifnum\hour<12 \global\amorpm={am}%
    \else\global\amorpm={pm}\advance\hour by-12 \fi
    \ifnum\hour=0 \hour=12 \fi
    \number\hour:\ifnum\minute<100\fi\number\minute\the\amorpm}}
\edef\militarytime{\number\hour:\ifnum\minute<100\fi\number\minute}
\def\draftlabel#1{{\@bsphack\if@filesw {\let\thepage\relax
  \xdef\@gtempa{\write\@auxout{\string
    \newlabel{#1}{{\@currentlabel}{\thepage}}}}}\@gtempa
    \if@nobreak \ifvmode\nobreak\fi\fi\fi\@esphack}
     \gdef\@eqnlabel{#1}}
\def\@eqnlabel{}
\def\@vacuum{}
\def\draftmarginnote#1{\marginpar{\raggedright\scriptsize\tt#1}}
\def\draft{\oddsidemargin -.5truein
        \def\@oddfoot{\sl preliminary draft \hfil
        \rm\thepage\hfil\sl\today\quad\militarytime}
        \let\@evenfoot\@oddfoot \overfullrule 3pt
        \let\label=\draftlabel
        \let\marginnote=\draftmarginnote

\def\@eqnnum{(\theequation)\rlap{\kern\marginparsep\tt\@eqnlabel}%
\global\let\@eqnlabel\@vacuum}  }
\def\preprint{\twocolumn\sloppy\flushbottom\parindent 1em
        \leftmargini 2em\leftmarginv .5em\leftmarginvi .5em
        \oddsidemargin -.5in    \evensidemargin -.5in
        \columnsep 15mm \footheight 0pt
        \textwidth 250mmin      \topmargin  -.4in
        \headheight 12pt \topskip .4in
        \textheight 175mm
        \footskip 0pt

\def\@oddhead{\thepage\hfil\addtocounter{page}{1}\thepage}
        \let\@evenhead\@oddhead \def\@oddfoot{} \def\@evenfoot{}
}
\def\titlepage{\@restonecolfalse\if@twocolumn\@restonecoltrue\onecolumn
     \else \newpage \fi \thispagestyle{empty}\c@page\z@
        \def\thefootnote{\fnsymbol{footnote}} }
\def\endtitlepage{\if@restonecol\twocolumn \else  \fi
        \def\thefootnote{\arabic{footnote}}
        \setcounter{footnote}{0}}  
\catcode`@=12 \relax

\def\ps@headings{\def\@oddfoot{}\def\@evenfoot{}
\def\@oddhead{\hbox{}\hfill
        \makebox[.5\textwidth]{\raggedright\ignorespaces --\thepage{}--
        \hfill }}
\def\@evenhead{\@oddhead}
\def\subsectionmark##1{\markboth{##1}{}}
}

\ps@headings

\relax

\def\firstpage#1#2#3#4#5#6{
\begin{document}
\newcommand{\newc}{\newcommand}
\newc{\ra}{\rightarrow}
\newc{\lra}{\leftrightarrow}
\newc{\beq}{\begin{equation}}
\newc{\eeq}{\end{equation}}
\newc{\bea}{\begin{eqnarray}}
\newc{\eea}{\end{eqnarray}}
\begin{titlepage}
\nopagebreak
\title{\begin{flushright}
        \vspace*{-0.8in}
{\normalsize CERN-TH/02-198}\\[-9mm]
{\normalsize hep-th/0208151}\\[-9mm]
\end{flushright}
\vfill
{#3}}
\author{\large #4 \\[1.0cm] #5}
\maketitle \vskip -7mm \nopagebreak
\begin{abstract}
{\noindent #6}
\end{abstract}
\vfill
\begin{flushleft}
{\normalsize CERN-TH/02-198}\\[-9mm]
\vspace*{.6cm}
{\normalsize August 2002}\\[-9mm]
\end{flushleft}
\thispagestyle{empty}
\end{titlepage}}

\def\simlt{\stackrel{<}{{}_\sim}}
\def\simgt{\stackrel{>}{{}_\sim}}
\date{}
\firstpage{3118}{IC/95/34} {\large\bf   Non-Collapsing Membrane
Instantons in Higher Dimensions }
{E.G. Floratos$^{\,a,b,d}$ and G.K. Leontaris$^{\,c,d}$}
{\normalsize\sl $^a$Institute of Nuclear Physics,  NRCS
Demokritos,
{} Athens, Greece\\[-3mm]
\normalsize\sl $^b$ Physics Department, University of Athens,
Athens, Greece\\[-3mm]
\normalsize\sl $^c$Theoretical Physics Division, Ioannina
University,
GR-45110 Ioannina, Greece\\[-3mm]
\normalsize\sl $^d$ CERN, Theory Division, 1211 Geneva 23,
Switzerland} {
 We introduce
a particular embedding of seven dimensional self-duality membrane
equations in $C^3\times R$ which breaks $G_2$ invariance down to
$SU(3)$. The world-volume membrane instantons define
$SU(3)$ special lagrangian submanifolds of $C^3$.
We discuss in detail solutions for spherical and toroidal topologies assuming
factorization of time.
 We show that the extra dimensions manifest themselves in the solutions
through the appearance of a non-zero conserved charge which prevents
the collapse of the membrane.  We find non-collapsing rotating membrane
instantons which contract from infinite size to a finite one and then they
bounce to infinity in finite time. Their motion is periodic.
These generalized complex Nahm equations, in the axially symmetric case,
lead to extensions of the continuous Toda equation to complex space.}
\newpage
\section{Introduction}

Some years ago, we introduced the notion of the self-duality for
supermembrane in $4+1$-dimensions and in the light-cone gauge. The
corresponding self-duality (s-d) equations proved to be an
integrable system with an infinite number of conservation laws and
particular  solutions were found~\cite{FL89,Ward90} which were
collapsing configurations of membrane instantons to point-like or
string-like objects. Similar covariant self-duality equations have
been introduced before for $2+1$-dimensions~\cite{BFS87} and later
generalized to $6+1$ dimensions in~\cite{Grabowski:cx}.

These objects, represent  world-volume instantons of the supermembrane.
 In the light cone gauge, the world-volume time and the target time are
 identical, so these configurations are space-time membrane instantons
 and they provide quantum mechanical tunnelling through the membrane
 self-interaction potential moving with velocities bigger than light.
Thus, they can travel infinite distances in finite time.

Their equations of motion, which are Nahm's type equations for the
area-preserving diffeomorphism  group of the membrane, lead for
axially symmetric configurations to continuous Toda equations
relating thus the membrane instantons with the self-dual Einstein
metrics with isometries~\cite{Bakas,Ward2}.

The basic ingredients for the study of covariant membrane
instantons in higher than ($4+1$)-dimensions, were contained in
the pioneering paper of ref~\cite{Grabowski:cx}, however this work
was until recently overlooked. The authors in~\cite{FCZ}
introduced higher dimensional self-duality equations for the
light-cone membranes as well as in the case of the quantized
Poisson (i.e. Moyal) bracket. The detailed properties of the
octonionic light-cone membrane instantons where studied
in~\cite{FL2,FLPT}, where  the invariance of the seven dimensional
equations under  the exceptional group $G_2$ was exploited. The
invariance under this group has as a consequence one remaining
supersymmetry consistent with the membrane background. In three
dimensions there are eight remaining supersymmetries~\cite{FL3}.

  In a parallel development, octonionic self-duality for seven and eight
  dimensional gravity was  proposed~\cite{Acharya:1996tw,Floratos:1998ba}
  and explicit seven and eight gravitational instantons which generalize
  four-dimensional ones (satisfying first order equations) were
  found~\cite{Floratos:1998ba,Bakas:1998rt}. These higher dimensional
  gravitational instantons where among the first few explicitly known
  self-dual metrics with exceptional holonomies $G_2$ and Spin(7) which
  were also lifted in 10 and 11-dimensional supergravity. Recently,
  exceptional holonomy higher dimensional instantons were studied for
  their r\^ole in string and M-theory and an important activity around
   this subject  has been created~\cite{Cvetic:2002kn}.

In this letter, we  introduce the complexified self-duality
equations of the membrane in seven dimensions and represent them
as generalized  Nahm equations. We show that the extra dimensions
manifest themselves in the solutions through the appearance of an
non-zero conserved charge which prevents the collapse of the
membrane. We integrate completely the three-dimensional complex
Nahm's equations for $S^2$ and $T^2$ topologies, assuming
factorization of time. We find periodic non-collapsing instantons.
Starting from infinite size they contract, with increasing angular
velocity, to a minimum size  and then they bounce back to infinity
in finite time.


\section{The Self-Duality Membrane equations in seven Dimensions.}

Choosing fixed values for the 8th and 9th membrane coordinates,
the seven-dimensional self-duality
equations~\cite{Grabowski:cx,FCZ,FL2} become
\begin{equation}
\dot{X}_i = \frac{1}{2} \Psi_{ijk}\{X_j,X_k\} \label{osce}
\end{equation}
where  $\Psi_{ijk}$ is the completely antisymmetric tensor that
defines the multiplications of octonions~\cite{gursey}. The Gauss
law results automatically by making use of the $\Psi_{ijk}$ cyclic
symmetry
\ba \{\dot{X}_i,X_i\}= 0
\ea
 The Euclidean equations of
motion  are obtained as follows
\ba
\ddot{X}_i& = & \frac 12\Psi_{ijk}\left(\{\dot{X}_j,X_k\}+ \{X_j,\dot{X}_k\}\right)\\
          & = & \{X_k,\{X_i,X_k\}\}
\ea
 where use has been made of the identity
 \ba
\Psi_{ijk}\Psi_{lmk}=\delta_{il}\delta_{jm}-\delta_{im}\delta_{jl}+\phi_{ijlm}
\label{I1} \ea
 and  of the cyclic property of the symbol
$\phi_{ijlm}$~\cite{gursey}.

At this point we would like to make a general remark on the nature
of the motion described by (\ref{osce}). These equations describe
the time evolution of the membrane instanton in flat space-times.
If the coordinates $X_i,\;(i=1,\dots,7)$ are periodic functions of
the membrane parameters $\sigma_{1,2}$, then integrating both
sides of the equations we find that all membrane instantons have
their center of mass pinched in a fixed point of space. This
implies spontaneous symmetry breaking of translational invariance.
If some of the flat space dimensions are compactified, then the
center of mass moves with the velocity determined by the cross
products of the winding numbers of the membrane in the
compactified dimensions. The cross  product is defined through the
tensor $\Psi_{ijk}$~\cite{FLPT}.

In what follows, we proceed to the complexification of the
self-duality equations. We embed the seven dimensional space $R^7$
into $C^3\times R$ in a very specific way which depends on the
particular definition of the octonionic structure constants used
in ref~\cite{FL2}  which assume  the following multiplication
table~\cite{gursey} 
\ba
\Psi_{ijk}=\left\{\begin{array}{ccccccc}1&2&4&3&6&5&7\\
                             2&4&3&6&5&7&1\\
                             3&6&5&7&1&2&4
 \end{array}\right.
\label{2.1}
\ea
Thus, if we define
\ba
z_1=X_1+iX_4,\;\;
z_2=X_2+iX_5,\;\;
z_3=X_3+iX_6,\;\;
a_0=X_7\nonumber
 \ea
then the self-duality equations become
\ba
D_t z_I\;=\;\frac{1}{2}\epsilon_{IJK}\{z_J^*,z_K^*\}, &\;\; D_t
a_0\;=\;\frac{\imath}2\{z_I,z_I^*\}\label{17}
\ea
where $I,J,K$ take the values $1,2,3$, whereas $D_t$ is  the
`covariant' derivative
\ba
D_t&=&\partial_t-i\{a_0,\;\;\}
\ea
These strikingly simple equations of self-duality, break the $G_2$
invariance down to $SU(3)$. The $SU(3)$ invariance comes from the
unique cross product existing in $C^3$ which is a remnant of the
octonionic cross product in seven dimensions. One consequence is
that, the three dimensional world-volume manifolds described by
(\ref{17}) are $SU(3)$ special lagrangian sub-manifolds of
$C^3$~\cite{Joyce}.

In the next section, we will consider the factorization of time
and the restriction to three complex dimensions of the above first
order equations. Before that, we would like to observe that it is
possible to generalize the connection of the three dimensional
self-duality equations with the continuous Toda
equations\cite{FL89,Ward2,Bakas}. This is possible if we consider
axially symmetric solutions of the above system. Indeed, the
axially symmetric Ansatz,
\ba
z_1=R(\sigma_2,t)\,\cos\,\sigma_1,\;\;\;
z_2=R(\sigma_2,t)\,\sin\,\sigma_1 ,\;\;\; z_3= z(\sigma_2,t)
\label{eqt}
\ea
where, $R,z$ complex functions, implies $\dot a_0=0$ for all times
and thus $a_0$ can be fixed to zero by an area preserving
transformation. For the equations (\ref{17}) we obtain
\ba
\dot R=R^*\, z^*_{\sigma_2},& \dot z=-R^*\,R^*_{\sigma_2}\label{toda1}.
\ea
and the index $\sigma_2$ refers to the derivative with respect to
$\sigma_2$. This system of equations has as integrability
condition the following non-linear equation which extends the
continuous Toda equation to three complex dimensions.
\ba
\frac{1}{R^*}\,{\partial^2_t R}-\frac{1}{R^{*2}}\,{\partial_t
R\,\partial_t R^*}+\frac
12{\partial^2_{\sigma_2}}\,R^{2}&=&0\label{Toda2}
\ea
This equation maybe relevant for the higher dimensional self-dual
gravity. Setting in (\ref{Toda2}) $R^2=e^{\Psi}$ we obtain the
form
\ba
\ddot\Psi+\frac 12 (\dot\Psi-\dot\Psi^*)
\,\dot\Psi+e^{-\frac{1}{2}(\dot\Psi-\dot\Psi^*)}\partial_{\sigma_2}^2e^{\Psi}=0
\ea
In the {\it real} three dimensional case~\cite{FL89}, we have
$R^*=R$ and $z=z^*$, while the continuous Toda equation for
$\Psi=\Psi^*$ reads~\cite{FL89,Bakas,Ward2},
\ba
{\partial^2_t\Psi}\,+\,{\partial_{\sigma_2}^2\,e^{\psi}}&=&0
\ea
 In the next section, among other things we find
the complete solution of the above (\ref{toda1}) system or of the
generalized continuous Toda equation (\ref{Toda2}) restricting the
functions $R, z$ so that $R(\sigma_2,t)=\sin\,\sigma_2\zeta(t)$,
$z(\sigma_2,t)=\cos\,\sigma_2\,\zeta_3(t)$.

\section{Membrane instantons in three complex dimensions}

As we show below, it is possible to extend the known  three
dimensional instanton solutions into six dimensions where, apart
from the radial expansion of the instanton, we observe rotational
motion in all of the three planes $(X_1,X_4), (X_2,X_5),
(X_3,X_6)$ of the six-dimensional space (which we choose to call
them I,II,III complex planes).

We assume factorization of time which will lead to a coherent
motion of all the membrane points. These solutions are analogous
(but for Euclidean time) to the real time solutions of second
order equations of motion for toroidal and spherical membranes
recently studied in\cite{Axenides:2002wi}
\ba
z_i&=& \zeta_i(t) f_i(\sigma_1,\sigma_2) \label{ans}
\ea
where  $f_i$ are three complex functions on the surface. First we
observe that the Poisson bracket $\{z_i,z_i^*\}=0$, if the
functions $f_i,f_i^*$ are functions of the same combination
$\sigma_1,\sigma_2$. From the equation for $a_0$ we find $\dot
a_0=0$ and therefore by an appropriate area preserving
transformation we may fix $a_0$ to be zero. So we are left with
the three complex Nahm's equations for $z_i$. We shall examine in
detail two topologies, -spherical ($S^2$) and toroidal ($T^2$).

Up to now only three-dimensional solutions of the self duality
equations are  known~\cite{FL89}. In order to factorize the time
dependence we choose for the case of $S^2$ the three functions
$f_i$ to be
\ba
f_1=\cos\phi\sin\theta,\; f_2=\sin\phi\sin\theta,\; f_3=\cos\theta
\ea
The three functions for the algebra $SU(2)$ under Poisson bracket
satisfy
\ba
\{f_i,f_j\}&=&-\,\epsilon_{ijk}f_k,\; i,j,k=1,2,3
\ea
{}For the three complex functions of time, we find the complex
Euler equations~\footnote{For the seven dimensional system in
another context, see also~\cite{UENO}.}
\ba
\dot \zeta_i&=&-\,\frac
12\,\epsilon_{ijk}^2\zeta_j^*\zeta_k^*\label{evolve}
\ea

In the case of $T^2$ we choose the following three functions
\ba
f_i&=&e^{\imath \vec{n_i}\cdot\vec\sigma},\, i=1,2,3,\\
\vec{n_i}&=&(n_{i1},n_{i2})\in Z^2
\ea
Now we observe that the factorization of time is implemented for
any three $\vec n_i$'s such that
\ba
\vec{n}_1+\vec{n}_2+\vec{n}_3=\vec 0
\ea
In this case, we obtain for the corresponding $\zeta_i(t)$
\ba
\dot \zeta_i&=&-\,n\frac 12\,\epsilon_{ijk}^2\zeta_j^*\zeta_k^*
\ea
where $n=n_{11}n_{22}-n_{12}n_{21}\in Z$.

In both cases ($S^2$ and $T^2$) the equations for the time
evolution are essentially the same. The $T^2$ case is obtained
from the equations of $S^2$ if we make the replacement $t\ra n t$
for $n$ integer. Therefore we only need to investigate the
equation (\ref{evolve}) which in component form is written \
\ba
\dot\zeta_1=-\zeta_2^*\zeta_3^*,\;\;\dot\zeta_2=-\zeta_3^*\zeta_1^*,\;\;
\dot\zeta_3=-\zeta_1^*\zeta_2^*\label{evolve1}
\ea
There is an obvious symmetry of the above system
\ba
\zeta_k&\ra&e^{\imath q_k}\zeta_k,\ea where $q_k, k=1,2,3$ are real and
$q_1+q_2+q_2=0$. This invariance leads to the
conservation of the three charges
\ba
Q_i&=& -\frac{\imath}2
\left(\dot\zeta_i\zeta_i^*-\dot\zeta_i^*\zeta_i\right),\;i=1,2,3
\label{Qis}
\ea
On the other hand, the equations of motion (\ref{evolve1}) imply
that all three charges  $Q_i$ are equal  to
\ba
Q_i\;\equiv\;{Q}&=&-\frac{\imath}2
\left(\zeta_1\zeta_2\zeta_3-\zeta_1^*\zeta_2^*\zeta_3^*\right)
\label{Qall}
\ea
There are  two additional constants of motion
in analogy with the Euler equations for the rigid body,
\ba
c_{ij}&=&|\zeta_i|^2-|\zeta_j|^2
\ea
where $c_{ij}$ are constants. In polar coordinates
\ba
\zeta_k&=&r_k \,e^{\imath\phi_k}
\ea
 we obtain
\ba
Q&=&r_1r_2r_3\,\sin(\phi_1+\phi_2+\phi_3)\label{red1}\\
\dot\phi_k&=&\frac{Q}{r_k^2}\label{40}\\
c_{ij}&=&r_i^2-r_j^2
\ea
Theen, equations (\ref{evolve1}) reduce to
\ba
\dot{r}_i&=&r_jr_k\cos\phi,\label{dri}
\ea
where $\phi=\phi_1+\phi_2+\phi_3$. For simplicity we define
$s_1=r_1^2$. We further combine (\ref{dri}) with (\ref{red1}) to
obtain the following differential equation
\ba
\dot{s}_1&=&-2\sqrt{s_1s_2s_3-Q^2}
\ea
After substitutions, the differential equation obtains a unique
form in the right-hand side for all $s_i$ which is
\ba
\dot{s}^2&=&4 \left[s (s-a) (s-b)-Q^2\right]\label{Weier0}
\ea
where $s_1=s$ $s_2=s-a$, $s_3=s-b$, where $a=c_{12},
b=c_{12}+c_{23}$.

If we define a new function of time  ${\cal
U}(t)=s(t)-\frac{a+b}3$, the differential equation becomes
\ba
\dot{\cal U}^{\,2}&=&4{\cal U}^{\,3}-g_2\,{\cal U}-g_3
\label{Weier}
\ea
which is recognized as the standard form of the Weierstrass
equation with solution the  (doubly periodic) Weierstrass function
$\,{\cal U}(t)={\cal P}(t;g_2,g_3)$, with
\ba
g_2&=& \frac{4 }{3}(a^2+b^2-a b)\label{g2,}\\
g_3&=&\frac{4}{27} (2 b^3+2 a^3-3 a^2b-3ab^2)+4Q^2\label{g3,}
\ea

Before we proceed to the analysis of the solution, we would
like to point out that there is an isomorphism between the
membrane and the matrix model solutions with factorization of time
for the spherical and toroidal topologies. This implies that the
above solutions have isomorphic matrix model instanton solutions
similar to those examined recently in ref\cite{Sfetsos:2001ku}.

\section{Analysis of the non-collapsing instanton solutions}

The equation of motion (\ref{Weier0}) for the membrane radii
($r_1^2\equiv s$, $r_2^2=s-a$ and $r_3^2=s-b$), is analogous to
the motion of a particle in a potential $V$. This is indeed the
motion of any point of the Euclidean membrane:
\ba
\dot{s}^2\;\equiv\;V(s,Q)\;=\; 4 [ s (s-a) (s-b)-Q^2] \label{psq}
\ea
where $a,b$ are positive constants. Without loss of generality we
may choose $a>b>0$. The left hand side of the  equation (\ref{psq}) is
always positive, thus the permitted regions for the variable $s$
are such that $V(s,Q)$ is also positive. These regions depend on
the values of $Q$. There is a limiting position of the qubic curve
when $Q^2=0$ ( $V(s,0)\equiv V_0(s)$). This is the upper curve
shown in figure (1) and the three real roots are $s_1=a$, $s_2=b$
and $s_3=0$. For all other values of $Q^2$ the curve is below the
limiting one and whenever there are three real roots, they are
$b>s_3>0$ $s_2<b, s_1>a$ respectively. $V(s,Q)$ possesses extrema
at the values
\ba
s_{max/min}&=&\frac 13 \left(a+b\mp\sqrt{a^2+b^2-ab}\right)
\label{minmax}
\ea
There is a critical value $Q^2=Q_c^2$ for which the $V(s,Q)$ has a
double root which is the $s_{max}$.
 $Q_c$ is determined as follows
\ba
Q_c^2&=& V_0(s_{max})
\ea
 In this case  we calculate the
maximum root to be
\ba
s_{1}^{c}&=& \frac 12 \left(a+b+2\sqrt{a^2+b^2-ab}\right)
\label{s1c}
\ea
The physical region is this case is beyond $s_1^c$.

\begin{figure}[h]
\begin{center}
\epsfig{file=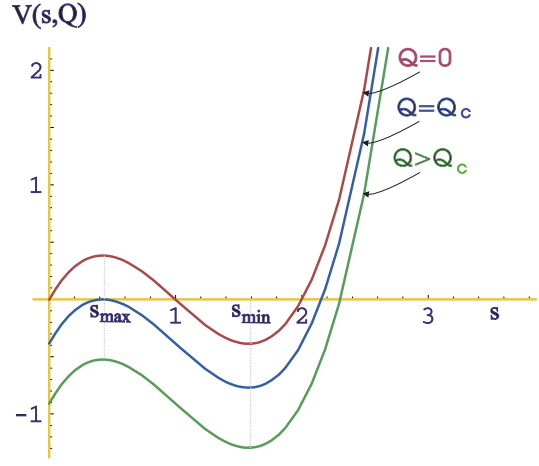,width=7cm}
\end{center}
\caption{{\it The `Potential' for three characteristic $Q$-values:
In all cases, the allowed region is beyond the highest root to the
infinity.}} \label{potfig}
\end{figure}

When $Q^2>Q_c^2$ there are two complex conjugate roots (the
maximum of $V$ is below the real axis) the physical region is
$s>s_1$ where $s_1$ is the real root. The three cases described
above are presented in figure (\ref{potfig}).
In what follows, we proceed in the detailed description of the dynamics
of the membrane instanton in the three cases discussed above ($Q=0$, $Q\ne 0$,
$Q=Q_c$).

 $\bullet$
When $Q=0$ it is possible to redefine the time-independent phases (\ref{40})
 to zero values and the self-duality equation reduces to the ones of the three
 dimensional case~\cite{FL89}. Because of the conservation laws we only need the equation
 for $s=r_1^2$ which reads
\ba
\dot{s}&=&-2\sqrt{s (s-a) (s-b)}\label{4dc}
\ea
We distinguish the following cases~\cite{FL89}:

\begin{itemize}
\item
 $a=b=0$. The solution is a spherical membrane with the
radius varied with time as
\ba
r&=&\frac{r_0}{1+r_0 (t-t_0)}\label{ssc}
\ea
There is a critical value $t_c=t_0-1/r_0$ where $r\ra\infty$,
whilst for $t\ra\infty$ the radius shrinks to zero.

\item
 $a=0, b\ne 0$. The equation becomes $\dot s=-2 s\sqrt{s-b}$
and the solution obtained is
\ba
r&=&r_0\frac{1+\frac{\sqrt{b}}{r_0}\,\tan\,\{\sqrt{b}(t-t_0)\}}{1-\frac{r_0}{\sqrt{b}}\,\tan\,\{\sqrt{b}(t-t_0)\}}
\ea
At $t_c=\frac{1}{\sqrt{b}}\tan^{-1}\frac{\sqrt{b}}{r_0}$ the
membrane has an infinite radius. On the contrary, when
$t_{in}=-\frac{1}{\sqrt{b}}\tan^{-1}\frac{\sqrt{b}}{r_0}$ the
configuration collapses to a string.

\item
 $a>b>0$. This is the most general four-dimensional case. The
s-d equation reads $$\dot s=-2\sqrt{s (s-a) (s-b)}$$
 In order to write it in a more familiar form,  we make the transformations
$x=\frac{\sqrt{a}}{r},\, k^2=\frac{b}{a}<1$, with $r>a$
and therefore $x<\frac{1}{\sqrt{a}}$. Then, separating variables
we have
\ba
t&=&\frac{1}{\sqrt{a}}\int_0^{\sin\phi}\frac{d\,x}{\sqrt{(1-x^2)
\,(1-k^2\,x^2)}},\;\;\,-\frac{\pi}2<\phi<\frac{\pi}2
\label{ellipform}
\ea
The right-hand side is the  elliptic integral~\cite{enderlyi}
${\cal F}(\phi,k)$, and the radius $r$ is given
\ba
r&=&\frac{\sqrt{a}}{sn\,\sqrt{a}\,t}
\label{solr}
\ea
Here, we assumed as initial condition $t=0$ and
$r_0=r(t=0)=\infty$. The positivity of $r$ restricts the $t$-range
in a half period of the elliptic $sn$. The real period is the
complete elliptic integral, i.e., when the upper limit of
(\ref{ellipform}) is equal to unity, $\sin\phi =1$,
\ba
\frac{T}2&=&\frac{1}{\sqrt{a}}\int_0^{1}\frac{d\,x}{\sqrt{(1-x^2)
\,(1-k^2\,x^2)}}\nonumber
\\
&=&\frac{1}{\sqrt{a}}\,K(k^2=\frac ba)
\label{elliperiod}
\ea
With the above initial conditions, at $t=0$  the volume of the
ellipsoidal  membrane is infinite, whereas at time $t=\frac T4$
reaches a minimum value with
$r_{min}=\frac{\sqrt{a}}{sn\,\sqrt{a}\,\frac T4}$  and the
membrane collapses to an elliptic disc.
\end{itemize}
It is worth mentioning that in three dimensions, assuming simple
factorization of time, we do not find non-collapsing membranes. On
the other hand we find all possible collapsed configurations for
the membrane, that is, points, strings and discs.

$\bullet$ Now we consider the  case $Q\ne 0$. This case
differentiates from the $Q=0$ case because it exists only in
dimensions higher than three ( see eq.\ref{40}).

As we shall see, a remarkable fact is that the dynamics of the membrane in
 higher dimensions is encoded in the higher dimensional angular momentum $Q$
  which, from the point of view of three dimensions, it behaves like a charge.

In the  case of spherical topology, there are three different
geometries, spherical, ellipsoidal with axial symmetry, and
anisotropic ellipsoidal ones. These three cases correspond to the
degeneracy of the roots of the polynomial in the right-hand side
of equation (\ref{Weier0}).

If the degeneracy $g$ is $g=3$, we have the spherically symmetric
membrane which from any initial condition it approaches the radius
equal to the largest real  root of the equation (\ref{Weier0}) in
finite time and it goes back to infinity.

If $g=2$, we have the axially symmetric ellipsoid which from an
initial configuration it decreases its volume until  a limiting
one which is determined also by the largest real root. The same
also happens to the anisotropic ellipsoidal membrane. The general
solution in terms of elliptic Jacobi functions or Weierstrass
function of the equation (\ref{Weier}), can be found in a similar
way with the case $Q^2=0$ in eq (\ref{4dc})
\ba
\frac{d s}{\sqrt{s(s-a)(s-b)-Q^2}}&=& \frac{d x}{\sqrt{x
(x-e_{13})(x-e_{23})}},
\ea
where $x=s-e_3$ and $e_{ij}=e_i-e_j$.

\begin{figure}[h]
\begin{center}
\epsfig{file=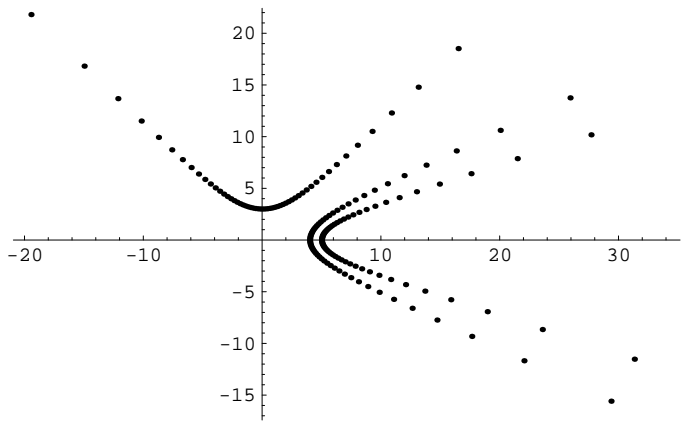,width=9cm}
\end{center}
\caption{{\it The points show the equidistant time-evolution of
the coordinates of the ellipsoidal membrane (\ref{solr}) in the
three planes $I=(X_1,X_4)$, $II=(X_2,X_5)$ and  $III=(X_3,X_6)$. All of
them reach a minimum size and go back to infinity. }}
\label{figvol}
\end{figure}

If the topology is toroidal, the radii $r_{1,2,3}$ of the torus
$T^3$ inside which the $T^2$ toroidal membrane is embedded, at any
moment of time they are equal to $\sqrt{s}, \sqrt{s-b},\sqrt{s-a}$
respectively. We note that in this case there are no
non-degenerate solutions below four  dimensions.

In the following, we discuss the $Q\ne 0$ spherical case
($a=b=0$). Integrating the equation we find the solution in terms
of the incomplete beta function~\cite{enderlyi}
\ba
t&=&\frac{1}{6 Q^{1/3}}\,Beta(\frac{Q^2}{r^6};\frac 16,\frac 12)
\ea
We assume here the following initial conditions: At $t=0$ the spherical
membrane has infinite volume and in finite time
$T=\frac{1}{6 Q^{1/3}}\,Beta(\frac 16,\frac 12)$ contracts  at the minimum
 permitted radius $r_0=Q^{1/3}$ and goes back to infinity. From the angular
 velocity equation (\ref{40}), at $t=0$ or infinite radius,  the angular
 velocity is zero and contracting it develops at the limiting time $T$
 angular velocity $\omega_T=\dot\phi=Q^{1/3}$.

The ellipsoidal cases follow similar pattern and we parametrize
the solution in terms of the Weierstrass function~\cite{enderlyi}:
\ba
s(t) &=&{\cal P}(t; g_2,g_3)+\frac{a+b}3\label{Weiers}
\ea
where $g_{2,3}$ are functions of $a,b$ given by (\ref{g2,},\ref{g3,}) above.

Due to the non-zero angular  momentum the brane  obtains a minimum
size given by the radii squared, $s_1, s_1-a, s_1-b$ where as
discussed in the beginning of this section, $s_1$ stands for the
largest root of $V(s,Q)$. At this minimum size, there are limiting
angular velocities given by
\ba
\omega_1\;=\;\frac{Q}{s_1},\;\;
\omega_2\;=\;\frac{Q}{s_1-a},\;\;\omega_3\;=\;\frac{Q}{s_1-b}.
\ea
In the solution (\ref{Weier}) we assume initial conditions
$s(0)=\infty$ and $s_1$ is given by $s_1={\cal
P}(\frac{T}2;g_2,g_3)+\frac{a+b}3$, where $T$ is the real period of the
Weierstrass function. In the  special case of $Q=Q^c$, we have
 a simple algebraic solution (similar to the $Q=0$ case), and
$s_1=s_1^c$ with $s_1^c$  given by (\ref{s1c}).

\section{Conclusions}

Breaking the $G_2$ invariance of the octonionic self-duality
equations for the membrane in seven dimensions down to $SU(3)$, we
found explicit solutions of non-collapsing rotating membrane
instantons which they have periodic motion starting at some
initial moment from infinite size, shrinking  down to a finite one
in a half period and then bouncing back to infinity. The r\^ole of
these instantons for the quantum mechanical vacuum of the membrane
depends on the period which is the inverse temperature in membrane
plasma of finite temperature. In the case of infinite period (zero
temperature) the membrane instantons collapse to point-, string-
and disc-like objects which represent the vacua of the quantum
mechanical membrane. Since up to now it is not known how to
quantize the supermembrane, we hope that the information we
provided in this work is a step towards this direction.

\vfill

{\it The authors would like to thank CERN-TH Group for kind
hospitality. GKL acknowledges partial support by European Union (contract
ref. HPRN-CT-2000-00152).}

 \end{document}